\documentclass[manuscript]{aastex}
\usepackage{graphicx}
\usepackage{rotating}
\usepackage{subfigure}
\usepackage{amsmath}
\usepackage{color}
\usepackage{eucal}
\usepackage{lineno}
\usepackage{natbib}
\linenumbers

\begin{document}

\title{A Comprehensive X-ray Absorption Model for Atomic Oxygen}

\author{T.~W.~Gorczyca\altaffilmark{1}, M.~A.~Bautista\altaffilmark{1}, 
M.~F. Hasoglu\altaffilmark{2},
J. Garc\'{i}a\altaffilmark{3},
E. Gatuzz\altaffilmark{4},
J.~S. Kaastra\altaffilmark{5,6}, T.~R. Kallman\altaffilmark{7}, S.~T. Manson\altaffilmark{8}, C. Mendoza\altaffilmark{1,4}, A.~J.~J. Raassen\altaffilmark{5,9}, C.~P. de Vries\altaffilmark{5}, and O.~Zatsarinny\altaffilmark{10} }
\altaffiltext{1}{Department of Physics, Western Michigan University, Kalamazoo, MI 49008, USA}

\altaffiltext{2}{Hasan Kalyoncu University, 27100 Sahinbey, Gaziantep, Turkey }

\altaffiltext{3}{Harvard-Smithsonian Center for Astrophysics, MS-6, 60 Garden
Street, Cambridge, MA 02138}

\altaffiltext{4}{Centro de F\'{i}sica, Instituto Venezolano de Investigaciones Cient\'{i}ficas, Caracas 1020, Venezuela}

\altaffiltext{5}{SRON Netherlands Institute for Space Research, Sorbonnelaan 2, 3584 CA Utrecht, the Netherlands}

\altaffiltext{6}{Sterrenkundig Instituut, Universiteit Utrecht, P.O. Box 80000, 3508 TA Utrecht, The Netherlands}

\altaffiltext{7}{NASA Goddard Space Flight Center, Greenbelt, MD 20771, USA}

\altaffiltext{8}{Department of Physics and Astronomy, Georgia State University, Atlanta, GA 30303, USA}

\altaffiltext{9}{Astronomical Institute Anton Pannekoek, 
Science Park 904, 1098 XH Amsterdam, University of Amsterdam, The Netherlands}

\altaffiltext{10}{Department of Physics and Astronomy, Drake University, Des Moines, IA 50311, USA}

\begin{abstract}
An analytical formula is developed to represent accurately the photoabsorption cross section of \ion{O}{1} for all energies of interest in X-ray spectral modeling. In the vicinity of the K~edge, a Rydberg series expression is used to fit
$R$-matrix results, including important orbital relaxation effects, that accurately predict the absorption oscillator strengths below threshold and merge consistently and continuously to the above-threshold cross section. Further minor adjustments are made to the threshold energies in order to reliably align the atomic Rydberg resonances after consideration of both experimental and observed line positions. At energies far below or above the K-edge region, the formulation is based on both outer- and inner-shell direct photoionization, including significant shake-up and shake-off processes that result in photoionization-excitation and double photoionization contributions to the total cross section. The ultimate purpose for developing a definitive model for oxygen absorption is to resolve standing discrepancies between the astronomically observed and laboratory measured line positions, and between the inferred atomic and molecular oxygen abundances in the interstellar medium from {\sc xstar} and {\sc spex} spectral models.

\end{abstract}



\section{Introduction}
\label{intro}

Atomic photoionization, an important astrophysical process, has been studied for more than a century since the seminal understanding of its energetics by \citet{einstein} and the first calculations of quantum mechanical cross sections \citep{bates}. Over the years, a plethora of experimental and theoretical investigations have managed an excellent grasp of its physics \citep{fanocooper, starace}, together with a remarkable quantitative description of the valence-shell photoionization of atoms and atomic ions \citep{opacity1,opacity2}. However, the quantitative model of inner-shell photoabsorption is less sound due to a variety of relaxation processes, namely Auger and X-ray emission, that must be taken into account in order to achieve acceptable accuracy, especially in the near-threshold region.

Inner-shell photoabsorption of metals with nuclear charge $7\leq Z \leq 28$ is directly accessible to modern X-ray observatories such as {\it Chandra} and {\it XMM-Newton}, and, hence, is of much interest in astronomy. Particularly prominent in the photoabsorption of the interstellar medium (ISM) are the K-shell features (lines and edges) of atomic oxygen, which is the most abundant metal and is critically important in the energetic and chemical evolution of the Universe \citep{stasinska}. At present, though, the unsatisfactory quantitative understanding of oxygen inner-shell photoabsorption is such that there exists various sets of cross sections, each one leading to different conclusions regarding the ionization and atomic-to-molecular fractions in the ISM along various Galactic lines of sight.

The first inner-shell photoabsorption cross sections of oxygen reported in the literature \citep{henke, verner93,verner95} were simple step-function fits to low-resolution solid-state data \citep{henke} or to theoretical calculations by \citet{reilman} using a central potential method. These results depict cross sections across inner-shell thresholds with unphysical discontinuous edges where even the threshold energies are poorly determined. These cross sections were used by \citet{schulz} in an early analysis of ISM absorption near the oxygen K~edge in {\it Chandra} X-ray binary-star spectra.

A later theoretical cross section, which took into account resonance effects, was computed by \citet{mclaughlin} using the {\it R}-matrix technique. However, this calculation failed to include the effects of orbital relaxation and spectator Auger damping, which causes blending of the resonances converging to the inner-shell thresholds, thus smearing the otherwise sharp K-shell edge. This cross section was used by both \citet{paerels} and \citet{takei} to analyze the ISM K-shell absorption of oxygen in the {\it Chandra} spectra towards X0614+091 and Cyg~X-2, respectively. All these studies found, after fitting the \ion{O}{1} K$\alpha$ line and edge, residual narrow absorption at $\approx$23.36 \AA\ and a broad edge feature at $\approx$22.9 \AA. In all cases, the residual absorption was most likely attributed to oxygen compounds although the narrow absorption feature could also be due to \ion{O}{2}.

By contrast, \citet{devries}, using the {\em XMM-Newton} Reflection Grating Spectrometer (RGS), found that the ISM oxygen K-shell edge observed in X-ray binaries and extragalactic sources was well described by the {\it R}-matrix cross section of \citet{mclaughlin}. A second {\it R}-matrix calculation of the \ion{O}{1} cross section was reported by \citet{o}, with full account of relaxation and Auger damping, in fairly good agreement with the laboratory measurements of \citet{stolte}. As a result, \citet{juett} analyzed the {\it Chandra} spectra available at the time using the cross sections by both \citet{mclaughlin} and \citet{o}, pointing out that the {\it R}-matrix cross sections and those by \citet{verner93,verner95} agreed to within $\approx$5\% well above threshold but with significant differences in the threshold region. These differences were such that, when using the more recent cross section to fit the spectrum, the previously found broad threshold residuals disappeared. Thus, it was concluded that the narrow absorption feature in the spectra, after subtracting the \ion{O}{1} contribution, was due to trace amounts of ionized oxygen rather than to molecular compounds. \citet{juett} also found that the discrepancies between the various calculations and experiments regarding the wavelengths of the \ion{O}{1} K$\alpha$ line and the K-shell threshold were considerably greater than the resolution of the astronomical spectra. Therefore, this energy dispersion is perhaps the main source of uncertainty left in the atomic cross sections.

Since then, \citet{Garcia} have reported {\it R}-matrix calculations for the whole oxygen isonuclear sequence which, in the case of \ion{O}{1}, agree to within $\approx 10\%$ with the near-threshold cross section of \citet{o}. Subsequently, \citet{Garciaoxygen} used these data sets to reanalyze seven {\it XMM-Newton} observations of the X-ray binary Sco~X-1, and by adjusting the absolute wavelength scale of the theoretical cross sections, found that the spectra were well fitted by \ion{O}{1} absorption alone with no conclusive evidence of contributions from any other source. On the other hand, a thorough study of {\it XMM-Newton} spectra towards the low-mass binary GS 1826-238 by \citet{pinto1} showed that the ISM was composed of a mixture of multi-phase gas, dust, and molecules; in the case of oxygen, its abundance was found to be 20--30\% higher than protosolar, and at least 10\% of its column density was in the form of molecules and dust grains. These findings have been confirmed by \citet{pinto2} in a more extensive survey of nine low-mass X-ray binaries where 15--25\% of the total amount of oxygen was found to be condensed in dust. Moreover, in a recent examination of several {\em Chandra} spectra towards the low-mass binary XTE J1817-330, \citet{gatuzz}---in order to fit the absorption lines from both the high- (\ion{O}{6}, \ion{O}{7}) and low-excitation (\ion{O}{1}, \ion{O}{2}, \ion{O}{3}) plasma components---were forced again to shift the photoionization cross sections of \citet{Garcia}, where the discrepancies pointed out by \citet{juett} regarding the observed and measured positions for the K$\alpha$ and K$\beta$ lines in  \ion{O}{1} still stood. \citet{gatuzz} report an oxygen abundance close to solar which in essence dissents from the conclusions reached by \citet{pinto1, pinto2}.

Laboratory measurements of the \ion{O}{1} K vacancy states \citep{caldwell, krause, menzel, stolte, mclaughlin2}, in particular the $1s2s^22p^5\ ^3P^o$ resonance, also show a bothersome scatter. This issue will be addressed more fully in Section~\ref{seclabexp} since this uncertainty, and that of the observations of interstellar oxygen X-ray spectra, are at the heart of the remaining issue of absolute energy normalization. The recent experiment of \cite{mclaughlin2} is similar to the earlier study by \citet{stolte}, but the entire resonance region is now covered in one continuous scan in photon energy as opposed to the previous piecemeal scans for the lower $n=2$ member and the higher $3 \le n\rightarrow\infty$ Rydberg series;  however, nearly identical energy positions are reported. The new theoretical results, on the other hand, are obtained from a 910-level $R$-matrix calculation that, upon close inspection, are essentially equivalent to those in \citet{o}.  Thus, there is nothing substantively new learned from this study other than a reconfirmation of the same resonance energy positions; this issue will be addressed more fully in Section~\ref{seclabexp}.

The ultimate purpose of the present study is to arrive at a consensus for the best description of the photoionization cross section for neutral atomic oxygen. To this end, it is advantageous to create a single photoabsorption model that is transparent to all atomic data users. This is most easily accomplished by formulating an analytical expression that includes all the essential features desired by modelers: accurate background cross sections (the ``shoulders''), line positions, widths, and oscillator strengths. We accomplish this by appealing to a combination of $R$-matrix computations, laboratory measurements, tabulated solid-state absorption data, independent-particle (IP) data, multi-configuration atomic structure calculations, and astronomically observed X-ray lines. Given a consistent photoabsorption cross section, we then plan to use the same atomic description in two different X-ray spectral modeling codes to render the ISM oxygen K features.

\section{Analytical Model Cross Section}
\label{Sec:Model}

We attempt to develop an analytical expression for the most reliable photoabsorption cross section possible. We begin with new $R$-matrix calculations, which are slight improvements over earlier results \citep{o} that were benchmarked favorably to experiment \citep{stolte}. These results, together with experimental measurements, allow for the most accurate representation of the strong $1s\rightarrow np$ resonances below the K~edge. At energies far below the K-edge region, where only outer-shell photoionization occurs, we use the formula of \citet{verner96} which is a simple fit to the IP results of \citet{reilman}, and is found to be in excellent agreement with the present $R$-matrix results. At higher energies, a fit to the tabulated data of \citet{henke} is used; these data are assessed to be the most accurate since important shake-up and shake-off processes are also accounted for (see Section~\ref{secrmat}). The resulting total photoabsorption cross section as a function of photon energy $E=h\nu$ is thus partitioned as
\begin{eqnarray}
\sigma_{PA}(E) & = &  \sigma_{2s,2p}(E) + \sigma^{\rm res}_{1s}(E) + \sigma^{\rm direct}_{1s}(E) \ .\label{eqtotal}
\end{eqnarray}
To illustrate this demarcation, we plot several data sets of the photoabsorption cross section in Fig.~\ref{figfittotal}, where
the three energy regions are all depicted.
For energies below $\approx 520$~eV, the cross section consists solely of the $\sigma_{2s,2p}(E)$ outer-shell photoionization contribution, whereas just below the K-edge region,
 the strong $1s\rightarrow np$ resonance absorption profiles dominate.
 At higher energies, the cross section is essentially due to direct
$1s\rightarrow \epsilon p$ photoionization and accompanying photoionization-excitation and double photoionization (which the present $1s$-photoionization $R$-matrix calculations do not include as discussed in Section~\ref{secrmat}). These three regions, and the precise theoretical modeling of each, are henceforth outlined.

\subsection{Outer-Shell Photoionization}

\begin{figure}[!hbtp]
\centering
\includegraphics[width=4.in,angle=-90.]{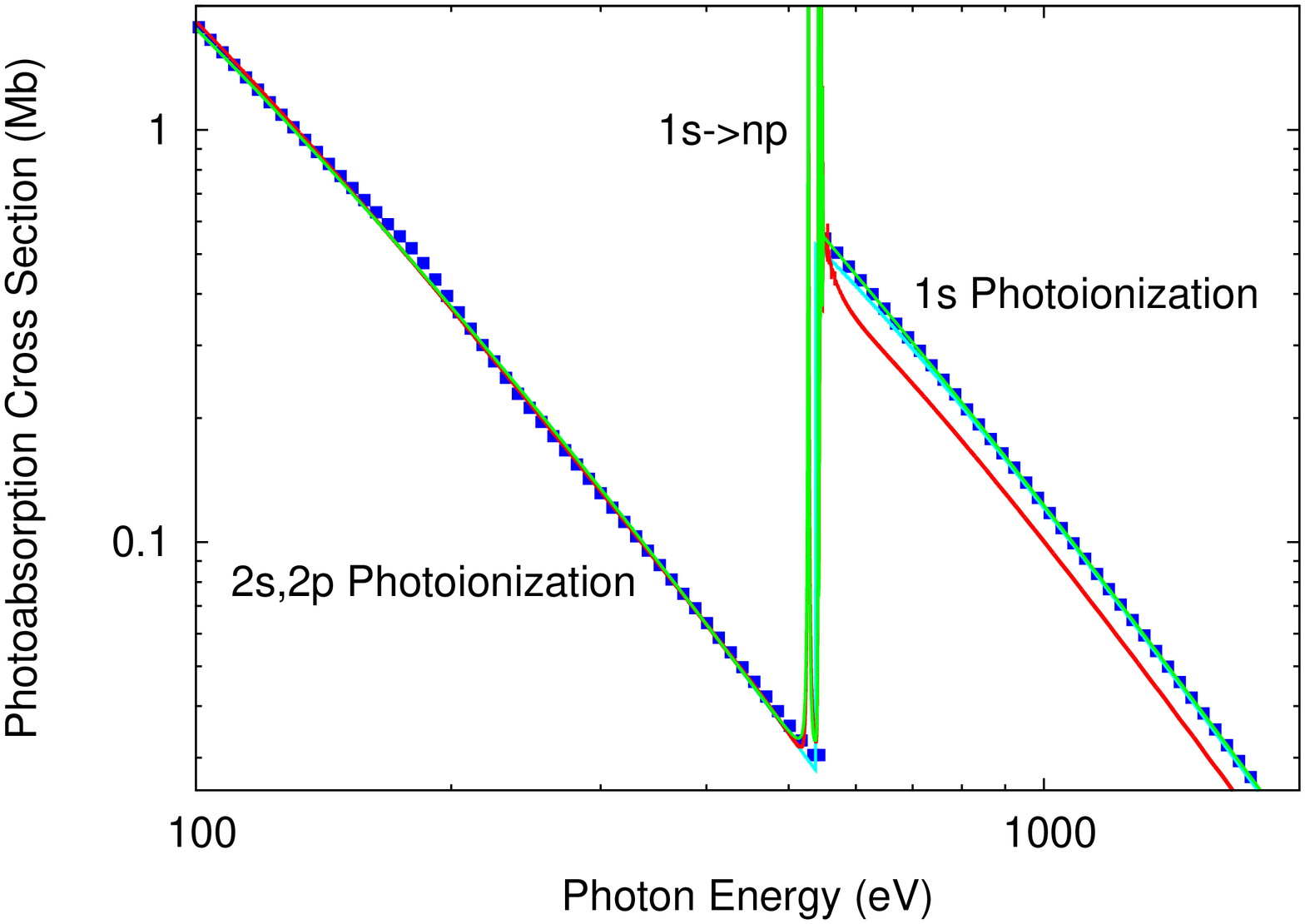}
\caption{\label{figfittotal}
A broad view depiction of the \ion{O}{1} photoabsorption cross section,
indicating where the outer shell ($\sigma_{2s,2p}$), inner-shell  ($\sigma^{direct}_{1s}$), and resonance ($\sigma^{res}_{1s}$) contributions are most important.
Shown are the present {\it R}-matrix (red curve), analytic formula (green curve),
IP fit of \citet{verner96} (cyan curve), and \citet{henke} data (blue squares).
The {\it R}-matrix results account for resonances but are missing two-electron contributions at higher energies (see text).  The fit incorporates all the correct physics.}
\end{figure}

For the outer-shell photoionization cross section, we have verified that the fit of \citet{verner96} is reliable, and their analytical formula
\begin{eqnarray}
\sigma_{2s,2p}(E) & = &  \sigma_0\left[\left(x-1\right)^2+y_w^2\right]y^{(p-11)/2}\left(1+\sqrt{y/y_a}\right)^{-p} \label{eqsig2s2p}
\end{eqnarray}
is therefore adopted, where $x  =  E/E_0-y_0$, $y  =  \sqrt{x^2+y_1^2}$, and the fitting parameters
$\sigma_0$,
$E_0$,
$y_a$,
$p$,
$y_w$,
$y_0$,
and $y_1$
are listed in Table~\ref{tablefit}.
The fit is seen in Fig.~\ref{figfittotal} to be in close agreement with the present $R$-matrix results and with the tabulated data of \citet{henke} at energies slightly above the $2s$ ionization threshold up through the region just below the $1s\rightarrow np$ resonances. At lower energies, important channel-coupling effects and prominent outer-shell resonance structure are found in the $R$-matrix results \citep[and modeled somewhat more crudely in][]{henke}. However, since we are not concerned with such low energies, the present fit is sufficient. It should be noted that, just below the resonance region at about 520~eV, there is coupling between the (open) $2s^{-1}\epsilon p$ and (closed)  $1s^{-1}n p$ channels that gives rise to a slight dip in the $2s$ $R$-matrix cross section (not observable on the scale of Fig.~\ref{figfittotal}) due to a transfer of oscillator strength \citep{manson1,manson2}. This small dip over such a narrow energy region is ignored in the present final model.

\subsection{Inner-Shell (High-Energy) Photoionization}

We find that the data of \citet{henke} above the $1s$ threshold can be accurately fitted with the expression
\begin{eqnarray}
\label{eqtail}
\sigma_{1s}^{\rm Henke}(E) & = & \sigma_{th}\left[1+\alpha_1\left(\frac{E_{th}}{E}\right)+\alpha_2\left(\frac{E_{th}}{E}\right)^2
\right]\left(\frac{E_{th}}{E}\right)^3\ ,
\end{eqnarray}
where the fit threshold position is chosen to be $E_{th}=544.544$~eV, giving a threshold cross section of $\sigma_{th}=1.07$~Mb and parameters $\alpha_1=-0.7227$ and $\alpha_2=0.2153$, which are needed for further fitting (see Table~\ref{tablefit}). We choose this functional form in our desire to derive an expression that matches continuously from below each of the two main \ion{O}{2} thresholds, and therefore, only the two parameters $\alpha_1$ and $\alpha_2$ are needed to get the correct shape of the above-threshold cross section.

\subsection{Resonance Region}

Our strategy in the $1s\rightarrow np$ resonance region is first to formulate an analytical fit (see Section~\ref{secanalyticalfit}) to the  results from new $R$-matrix calculations
(detailed in Section~\ref{secrmat}).  The fit parameters are then adjusted slightly to match the experimental resonance positions and oscillator strengths, highlighting both the best assessment of the absolute energy scales as determined by line observations and the smooth consistent merging to the above-threshold and high-energy cross section.

\subsubsection{{\it R}-matrix Calculations}
\label{secrmat}

The present $R$-matrix approach is based closely on the earlier work of \citet{o}, the essential difference being in the present removal of pseudo-resonances. Further improvements include, firstly, a new computation of the spectator Auger widths using a resonance time-delay matrix analysis \citep{timedelay} within an independent $R$-matrix calculation for the ${\rm e}^- -$\ion{O}{2} scattering. A second improvement involves a smooth turn-off of the spectator Auger damping $E\rightarrow E+i\Gamma/2$ \citep{ar,o} as the effective quantum number approaches the orbital angular momentum ($\nu\downarrow l$) to avoid the discontinuity previously seen when the quantum defect approach is abruptly turned off.
Here we allow the width to vanish continuously ($\Gamma\rightarrow 0$) in this limit before the quantum defect channel is ``closed" off \citep{deep}.

As in earlier work \citep{o}, we also emphasize the critical importance of accounting for orbital relaxation following inner-shell photoionization: the $2s_r$ and $2p_r$ ``relaxed" orbitals in the final $1s2s_r^22p_r^4$ \ion{O}{2} vacancy state distinctively differ from those in the initial \ion{O}{1} $1s^22s_g^22p_g^4$  ground state due to the doubling of the  effective charge seen in the \ion{O}{2} state. As a result, the computation of the direct cross section involves an overlap amplitude
factor proportional to the $\langle 2s_g\vert 2s_r\rangle$ and $\langle 2p_g\vert 2p_r\rangle$ orbital overlap integrals, and it is therefore imperative to account for this orbital difference.  Within an orthonormal basis methodology, such as the
$R$-matrix approach we use here \citep{burke,rmatcpc}, the only way to accomplish this is by introducing {\em pseudo-orbitals} (e.g. $\overline{3s}$, $\overline{3p}$, etc.) such that the relaxed excited state can be described in terms of the ground state and the pseudo-orbitals via
\begin{eqnarray}
1s2s_r^22p_r^4 & = & c_1 1s2s_g^22p_g^4 + c_2 1s2s_g^22p_g^3\overline{3p} + c_3 1s2s_g2p_g^4\overline{3s} + ... \ .
\end{eqnarray}
This procedure takes care of the relaxation effect, and in the present case, the reduction factor can be independently computed from simple Multi-Configuration Hartree--Fock (MCHF) calculations \citep{ffmchf} to be $\vert c_1\vert^2=0.80$. Therefore, there is an analytically predicted reduction effect by a factor of 0.80 due to relaxation, and the remaining 20\% of the oscillator strength goes into photoionization-excitation and double photoionization. An excellent discussion of these various contributions is given in the early experimental study of \ion{Ne}{1} photoionization by \citet{wkrause} (see, especially, their Fig.~8).

\begin{figure}[!hbtp]
\centering
\includegraphics[width=1.8in,height=5.in,angle=-90.]{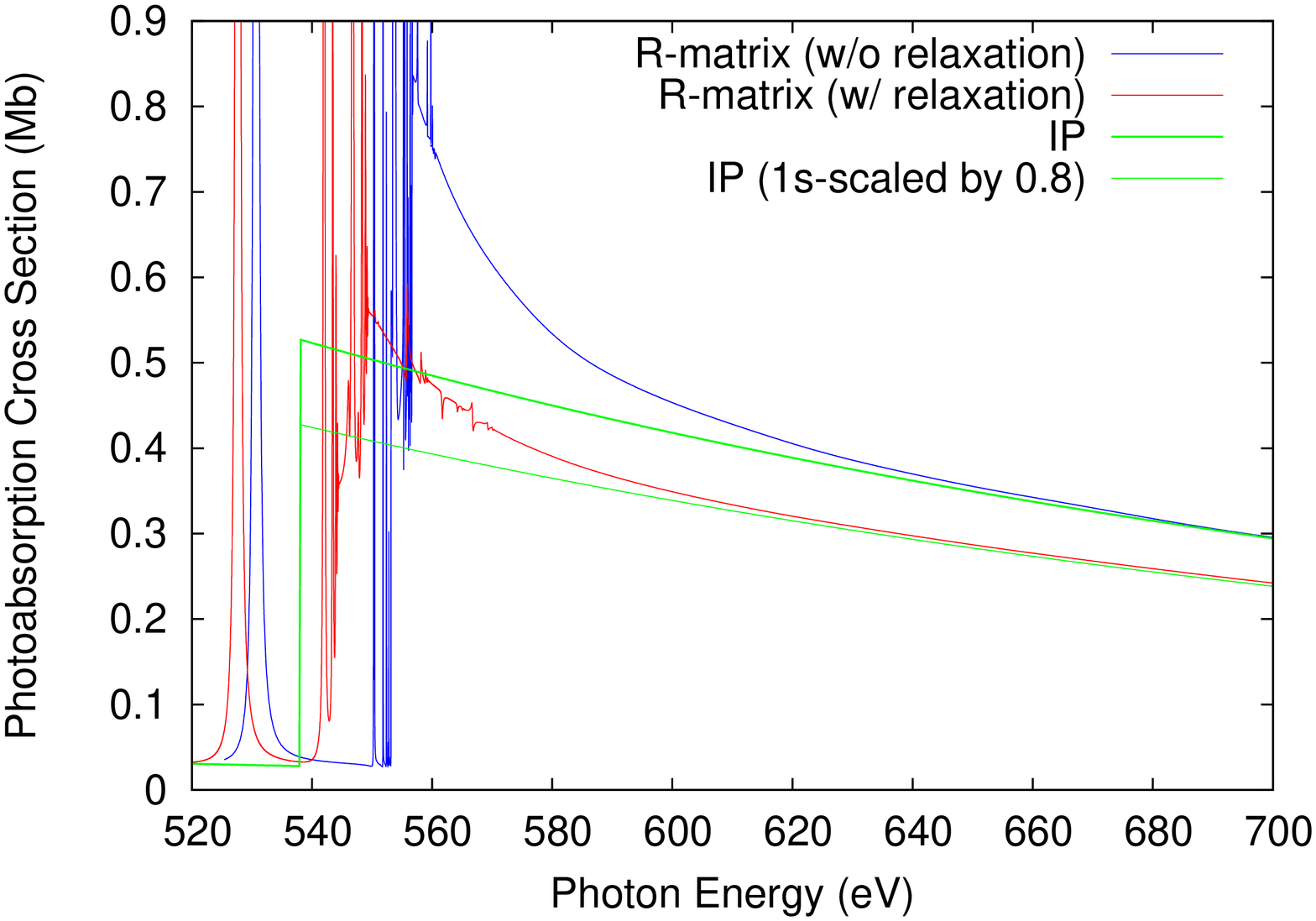}
\includegraphics[width=1.8in,height=5.in,angle=-90.]{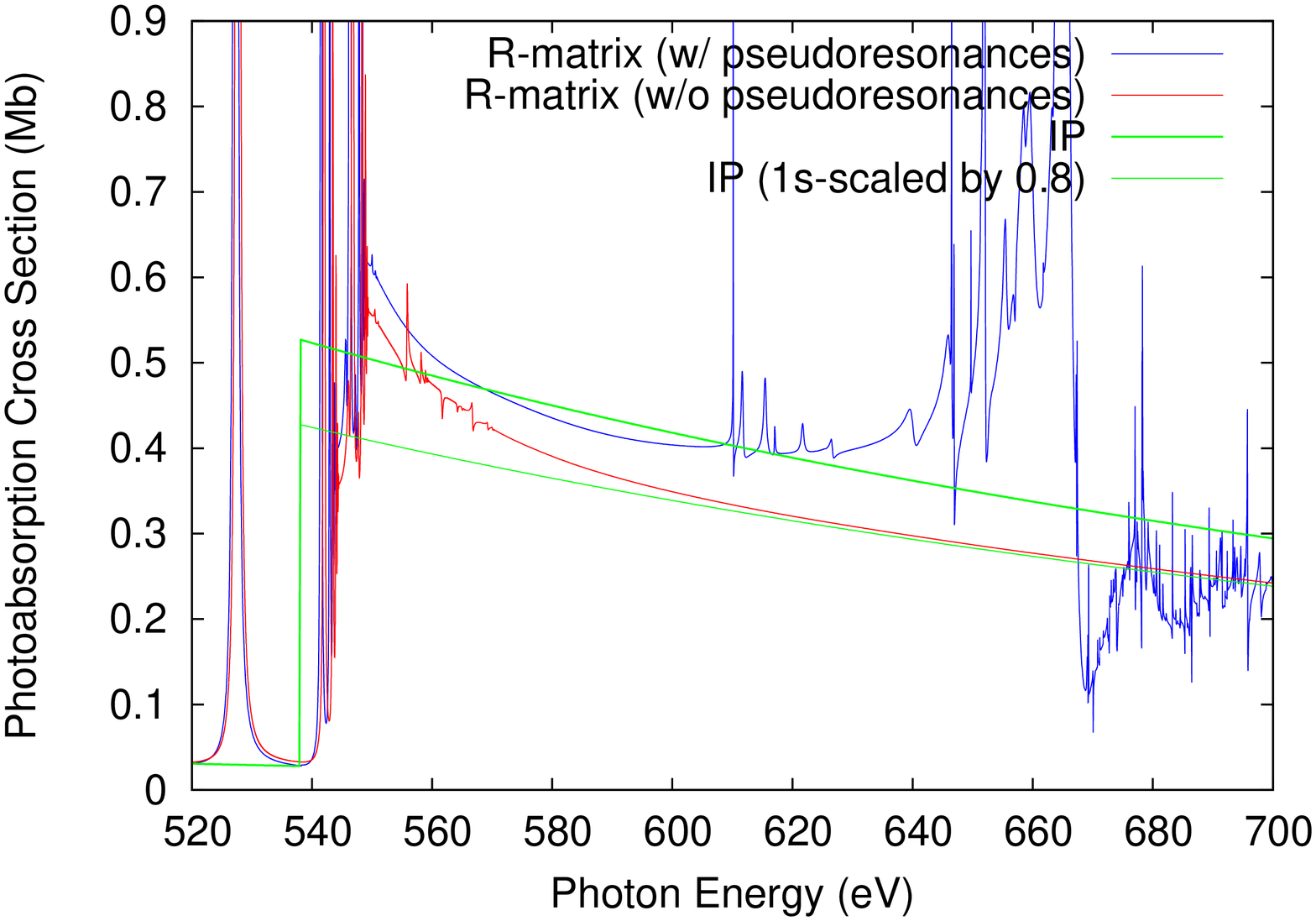}
\includegraphics[width=1.8in,height=5.in,angle=-90.]{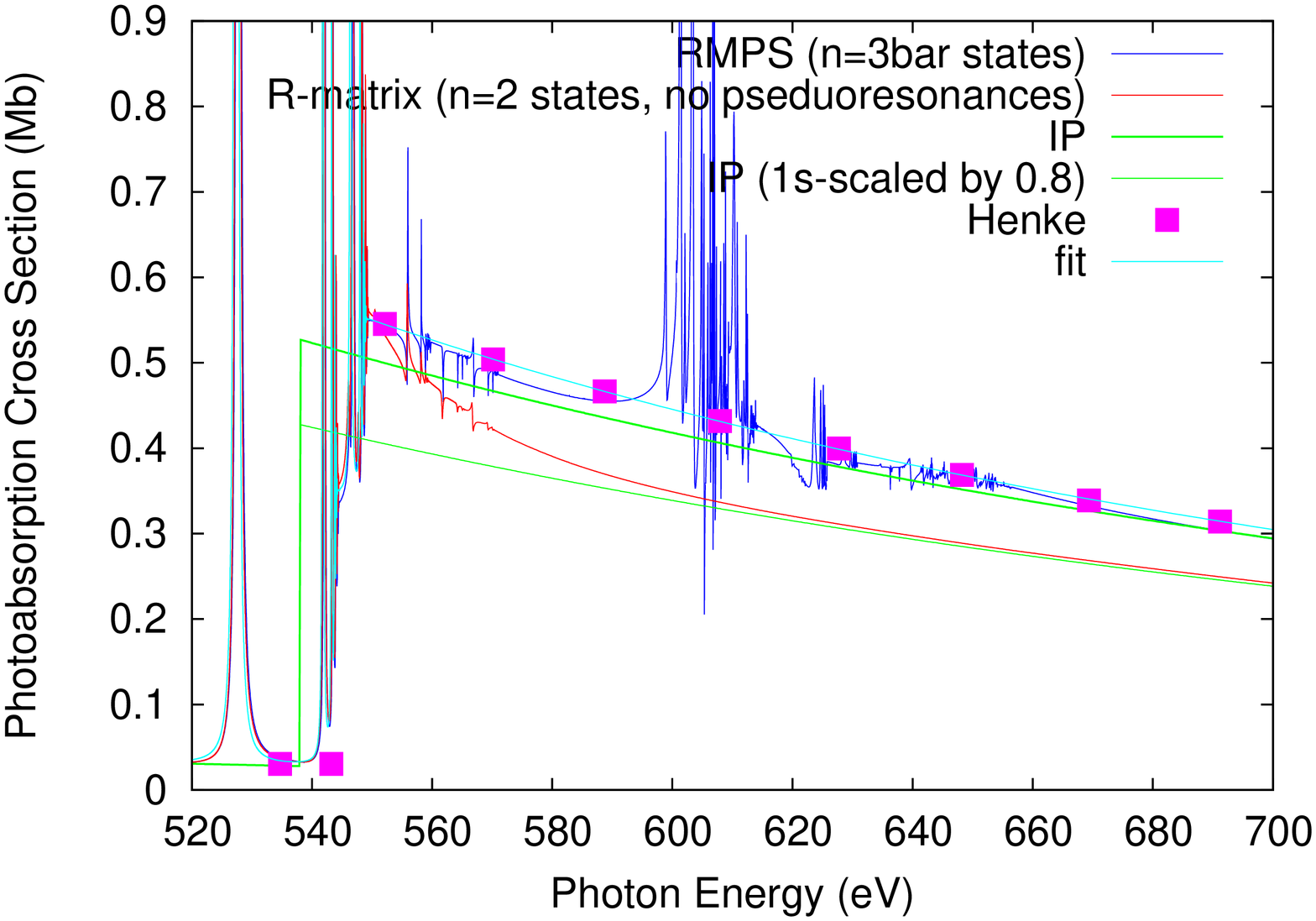}
\caption{{\it R}-matrix photoabsorption cross sections computed with and without relaxation effects and/or pseudoresonance elimination.  The red curve shows the definitive {\it R}-matrix calculation, which includes relaxation effects via the use of pseudoorbitals.  The blue curve in the upper plot shows
results without relaxation included, giving a gross overestimate of the threshold energy position
and cross section.  The blue curve in the middle plot shows the results when using pseudoorbitals
but not using the pseudoresonance elimination method \citep{pseudo}, giving large, unphysical
features which permeate the threshold region and below.  The two green curves show the
 IP asymptote and that reduced by 80\% due to relaxation effects.
 The blue curve in the lower plot shows the results when including the additional $1s2s^22p^3\overline{3p}$ and $1s2s2p^4\overline{3s}$ target pseudostates to give an approximate representation of the photoionization-excitation and double photoionization channels, while eliminating all additional pseudoresonances that are not associated with these pseudochannels.
\label{figrmat}
}
\end{figure}

The effect of relaxation can also be seen by comparing the present $R$-matrix cross section with one where relaxation is not taken into account (Fig.~\ref{figrmat}).  The present, final $R$-matrix cross section is seen to approach asymptotically the IP fit cross section of \citet{verner96}, namely the scaled $1s$ cross section after the latter has had the $1s$ contribution multiplied by a factor of 0.80 -- the same overlap factor we compute from an independent MCHF calculation thus independently confirming the 20\% reduction effect. It should be noted that the original IP calculations \citep{reilman}, upon which \citet{verner96} based their fit, did not include relaxation effects. Consequently, their asymptotic value reflects the {\em total} photoabsorption cross section, which includes shake-up and shake-off processes in addition to the direct $1s$ photoionization (without secondary excitation or ionization).

However, as also seen in Fig.~\ref{figrmat}, even though the unrelaxed orbital results approach the full IP cross section \citep{verner96} asymptotically, they grossly overestimate the correct cross section just above the K-shell threshold; here only the (relaxed) direct photoionization is energetically allowed.  This overvalue carries over below threshold, leading to an unphysically enhanced resonance oscillator strengths. Moreover, note that the threshold energy {\em position} is also overestimated---by more than 10 eV---due to the inaccurate representation of the $1s2s^22p^4$ inner-shell vacancy state.

It is therefore critical to account for relaxation effects; this is accomplished in the present theoretical methodology by including additional pseudo-orbitals in the atomic orbital basis set. However, without proper care, this procedure can lead to spurious {\em pseudo-resonance} structure \citep{pseudo} as is also shown in Fig.~\ref{figrmat}. The present $R$-matrix results, which were computed with pseudo-orbitals and with a proper elimination of pseudo-resonances \citep{pseudo}, are compared to similar $R$-matrix results without such an elimination. It is seen that the latter cross section exhibits large, spurious pseudo-resonance structure at higher energies.  Furthermore, these broad, unphysical resonance features are seen to permeate even down to the threshold region, resulting in an overestimate of the near threshold cross section (and the resonance absorption oscillator strengths below threshold).  By applying the pseudo-resonance elimination method \citep{pseudo}, the cross section turns out to be smooth throughout, and provides the most reliable resonance oscillator strengths as discussed in Section~\ref{secrmat}.  It is to be noted that the earlier $R$-matrix calculations did not use a pseudo-resonance elimination method and, therefore, overestimated the resonance absorption oscillator strengths; this was pointed out by \citet{o} and also depicted in their Fig.~2.

We can improve the asymptotic situation somewhat by including the orthogonal compliments to the $1s2s^22p^4$ O$^+$ (relaxed) target states, namely the additional {\em pseudo-states} which are composed of the $1s2s^22p^3\overline{3p}$ and  $1s2s2p^4\overline{3s}$ configurations (with smaller $\approx$20\% mixing of the $1s2s^22p^4$ configurations).  This resultant, so-called $R$-matrix with Pseudo-States (RMPS) method \citep{burke}, as implemented in the present codes following the developments of \citet{hermps}, gives a somewhat crude, approximate description of the photoionization-excitation and double photoionization channels, and importantly, leads to the correct high-energy photoionization asymptote, as seen in  the lower panel of Fig.~\ref{figrmat}. We note that the implementation of R-matrix methods on modern massively parallel machines \citep{parallel} will allow for
a much larger, converged RMPS treatment of the problem. 

\begin{figure}[!hbtp]
\centering
\includegraphics[width=4.in,angle=-90.]{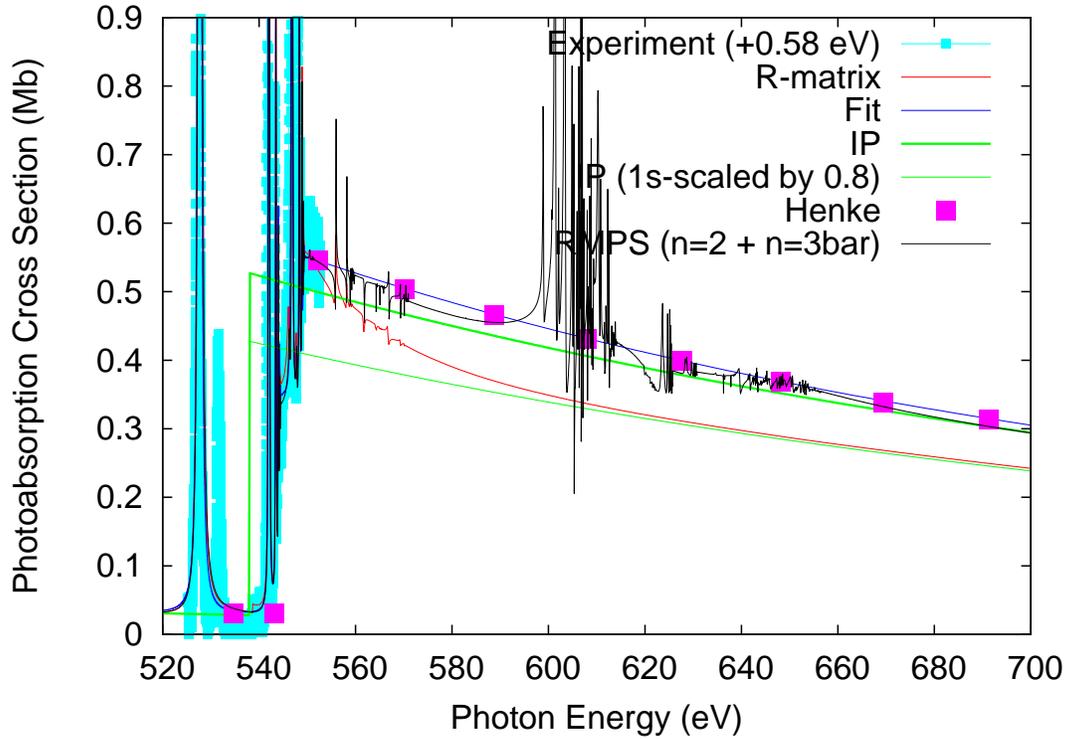}
\caption{\label{highe}
Near- and above-threshold cross section:
The red curve and black curve show the definitive {\it R}-matrix and RMPS calculations, respectively.
The two green curves show the
 IP asymptote and that reduced by 80\% due to relaxation effects.
 The final fit formula is shown as the blue curve, the experimental results of \citet{stolte}, shifted by +0.58 eV, are given as the cyan data points, and the solid-state results of \citet{henke} are given as the magenta squares.
}
\end{figure}

The findings thus far regarding the above-threshold cross section are summarized in Fig.~\ref{highe}: asymptotically, the present $R$-matrix cross section approaches the IP results after the (dominant) $1s$ contribution has been scaled by a factor of 0.8 to account for relaxation, whereas the RMPS values show the correct asymptote but are still plagued by pseudo-resonances.  The fit of \citet{verner96} at threshold is an extrapolation of the high-energy IP cross section of \citet{reilman}, and therefore, does not include the correct threshold rise as is seen in the $R$-matrix results. On the other hand, the data of \citet{henke}, which are based on solid-state measurements, show the correct threshold and asymptotic cross sections and are devoid of pseudo-resonance structure; therefore, we choose this continuous data as the best representation of the cross section for energies above threshold. Note that the $R$-matrix and RMPS cross sections just above threshold coincide with the data by
\citet{henke}. Lastly, note that the measurements of \citet{stolte} are consistent with the $R$-matrix results throughout, as we further address in Section~\ref{secfinalfit}.

\subsubsection{Analytical Fit to the $1s\rightarrow np$ Resonance Region}
\label{secanalyticalfit}

The formula to fit the single-resonance photoabsorption cross section---parameterized by an absorption oscillator strength $f$, a resonance position $E_r$, and a width $\Gamma$---is given by \citep[see][Eq.~71.19]{bethesalpeter}
\begin{eqnarray}
\sigma_{PA}(E) & = &  \frac{\pi (k_ee^2)h}{mc}\frac{df}{dE}\ ,\label{eqsigpa1}
\end{eqnarray}
where the oscillator strength per unit energy for an isolated resonance takes the form
 \begin{eqnarray}
\frac{df}{dE} & = &  f\ \frac{\Gamma/2\pi}{(E-E_r)^2+(\Gamma/2)^2}\ ; \label{egdfde}
\end{eqnarray}
i.e. it is equal to the discrete oscillator strength $f$ times an energy-normalized Lorentzian
\begin{eqnarray}
\int dE \frac{\Gamma/2\pi}{(E-E_r)^2+(\Gamma/2)^2} =  1\ .
\end{eqnarray}
Therefore, the photoabsorption profile can be characterized as
\begin{eqnarray}
\sigma_{PA}(E) & = &  \beta f\frac{\Gamma/2\pi}{(E-E_r)^2+(\Gamma/2)^2} \label{eq1b}
\end{eqnarray}
with
\begin{eqnarray}
\beta & = & \frac{\pi k_ee^2h}{mc} =  109.7626\,{\rm Mb\,eV} \ .\nonumber
\end{eqnarray}

For an entire Rydberg series---characterized for each member by a principal quantum number $n$, resonance positions $E_n$, width $\Gamma$ ($n$-independent for inner-shell spectator Auger decay), and the oscillator strengths $f_n$---can be parameterized by a quantum defect $\mu$, a threshold energy $E^{th}$, and an $n$-independent  ``strength'' $f_0$:
\begin{eqnarray}
E_n & \approx & E_{th}-\frac{Z^2E_{au}}{2(n-\mu)^2} \ , \nonumber \\
f_n & \approx & \frac{f_0}{(n-\mu)^3} \ ,\label{eq3}
\end{eqnarray}
where $E_{au}=27.211\,{\rm eV}$.
Noteworthily, this discrete expression carries over to an analytic above-threshold cross section
\begin{eqnarray}
\lim_{E\downarrow E_{th}}\sigma_{PA}(E) & = &  \beta\frac{f_0}{Z^2E_{au}}\ ,
\end{eqnarray}
which must be considered when developing a consistent, continuous formulation through threshold.

The above formula, based on quantum defect theoretical considerations, is precise as $n\rightarrow \infty$ but deficient for the lower resonances ($n=2,3$). The lowest members are more appropriately modeled by using separate (energy-dependent) quantum defects and oscillator strengths.

For multiple Rydberg series, if the interaction between them is neglected, the contribution from each series can be considered separately. We apply this approach to the two dominant photoabsorption series in oxygen, namely $1s2s^22p^4(^4P)np$ and $1s2s^22p^4(^2P)np$ (labeled, respectively, by the indices $i_s=1$ and $i_s=2$), giving a two-series resonance cross section parameterized as
 \begin{eqnarray}
\sigma^{res}_{1s}(E) & = & \beta\,\sum_{i_s=1}^2\left[\sum_{n=n_{min}=i_s+1}^\infty
\frac{f^{i_s}_{0,n}}{ (n-\mu_n)^3}
\frac{\Gamma^{i_s}_n/2\pi}
{ \left( E-(E^{i_s}_{th}+Z^2E_{au}/(n-\mu_n)^2) \right)^2
+\left(\Gamma^{i_s}_n/2\right)^2}\right.\nonumber \\
& & +  \left.
\frac{f^{i_s}_{0,\infty}}{Z^2E_{au}}
\left(\frac{1}{2}-\frac{1}{\pi}\arctan{\left(\frac{E^{i_s}_{th}-E}{\Gamma/2}\right)}\right)
\right]\ .\label{eqfit}
 \end{eqnarray}
The last term ensures that, since the below-threshold contribution has effectively been Auger broadened, i.e. convoluted with a Lorentzian of width $\Gamma$ within each resonance interval of energy region $\Delta E \sim E_{au}/(n-\mu)^3$, the step function due to the above-threshold continuum photoionization is likewise convoluted near threshold:
\begin{eqnarray}
\sigma(E\approx E_{th})=\int_{-\infty}^{\infty}dE' \frac{\Gamma/2\pi}{(E-E')^2+(\Gamma/2)^2}\left[\frac{f_{0,\infty}}{Z^2E_{au}}\theta(E-E^{th})\right]\ ,
\end{eqnarray}
where $\theta(E-E^{th})$ denotes the Heaviside step function at threshold. As the energy is increased above threshold, this expression is continuously extended to have the correct asymptotic tail as determined from our fit to the \citet{henke} data in Eq.~\ref{eqtail}, taking instead the form above threshold
 \begin{eqnarray}
\sigma^{\rm direct}_{1s}(E) & = & \beta\,\sum_{i_s=1}^2
\frac{f^{i_s}_{0,\infty}}{Z^2E_{au}}
\left[\frac{1}{2}-\frac{1}{\pi}\arctan{\left(\frac{E^{i_s}_{th}-E}{\Gamma/2}\right)}\right] \nonumber \\
& & \times
\frac{\left[1+\alpha_1\left(\frac{E^{i_s}_{th}}{E}\right)+\alpha_2\left(\frac{E^{i_s}_{th}}{E}\right)^2 \right]}
{[1+\alpha_1+\alpha_2]}\left(\frac{E^{i_s}_{th}}{E}\right)^3
\ \ .\label{eqfit2}
 \end{eqnarray}

Of particular note in our $R$-matrix calculations is that the oscillator strength (cross section) is found to be partitioned into the two dominant series by the fractions of 3/5 and 2/5 for the $1s2s^22p^4(^4P)np$ and $1s2s^22p^4(^2P)np$ series, respectively, instead of the statistical weighting of 2/3 and 1/3, due to channel coupling in the threshold region. As a result, the net oscillator strength density above threshold, which we find to be $f_{0,\infty}=0.132$, is partitioned as $f^1_{0,\infty}=0.6f_{0,\infty}$ and $f^2_{0,\infty}=0.4f_{0,\infty}$

Our strategy is first to fit this expression to our present $R$-matrix results to get a good representation of the oscillator strengths and quantum defects, using the threshold energies and widths from the $R$-matrix runs. Then, the threshold energies are slightly adjusted and the analytical (Lorentzian) fit is further convoluted with the experimental (Gaussian) width to obtain a good fit to the experimental resonance spectrum. But first a brief digression to address resonance energy positions.

\subsubsection{Resonance Energy Positions from Astronomical Observations}\label{sec_obs}

Oxygen K-shell photoabsorption in the ISM has been observed with both the {\it Chandra} and {\it XMM-Newton} satellite-borne observatories. The High Energy Transmission Grating Spectrometer (HETGS) in combination with the Advanced CCD Imaging Spectrometer (ACIS) of the former perhaps provide the best spectral resolution with adequate sensitivity. It is exemplified by the study of \cite{juett} using the Medium Energy Gratings \citep[MEG, resolving power of 0.023~{\AA} FWHM and an absolute wavelength accuracy of 0.011~{\AA},][]{can05} from the HETGS to observe six Galactic X-ray sources. For the present work, we have carried out a reanalysis of these observations in an attempt to improve the positions of the O~{\sc i} K$\alpha$ ($1s\rightarrow 2p$) and K$\beta$ ($1s-3p$) resonances; the XTE~J1817-330 source, previously treated by \cite{gatuzz}, has also been included and, when possible, additional spectra for the sources considered by \cite{juett}.

Observational specifications for the seven low-mass X-ray binaries used in this analysis are listed in Table~\ref{tab1}. The observations were taken in continuous clocking mode (CC) or time exposure mode (TE). In CC mode, the temporal resolution is increased in order to minimize the pileup effect \citep{cac08b}. In TE mode, the ACIS instrument reads the collected photons periodically.  All the spectrum files, response files (RMF), auxiliary response files (ARF), and background files were taken from the {\it Chandra} Grating-Data Archive and Catalog TGCat\footnote{http://tgcat.mit.edu/}. We have
used the {\sc isis}\footnote{http://space.mit.edu/cxc/isis/} package (version
1.6.2-18) for spectral fitting.

We have fitted all the observations for each source simultaneously using a simple {\tt powerlaw+gaussians} model in the oxygen-edge region (21--24~{\AA}). The power-law parameters were taken as independent free parameters for each observation. Cash statistics was applied due to the low signal-to-noise ratio in these spectra, which requires a minimal grouping of the spectra of at least one count per spectral bin \citep[see][]{hum09, bal12}. Table~\ref{tab2} shows the \ion{O}{1} K$\alpha$ and K$\beta$ absorption line positions obtained from these fits. Firstly, we list the mean position values excluding XTE~J1817-330 in order to compare with those originally reported by \citet{juett}. When this source is taken into account, we note a slight decrease of the mean-value error bars. In the case of Cygnus~X-1, the values in \citet{juett} correspond to the average of ObsID 3407 and ObsID 3742. From the confidence intervals it may be appreciated that the present effort results are in excellent agreement with the line positions estimated by \citet{juett}, but with improved statistics.

The transition energy of the \ion{O}{1} K$\alpha$ ($1s\rightarrow 2p$) line has also been estimated using the well-exposed {\it XMM-Newton} RGS spectrum of Mrk 421 \citep{kaastra2006}. This spectrum shows strong interstellar K$\alpha$ absorption lines from \ion{O}{1} at $23.5138 \pm 0.0022$~\AA\ and \ion{O}{7} at $21.6027\pm 0.0021$~\AA. Since accurate theoretical values \citep{drake, cann} and high-precision laboratory measurements \citep{engstrom} have been reported for the latter at 21.6015~\AA\ and $21.60195\pm 0.0003$~\AA, respectively, a small offset on the wavelength scale for this data set of 0.8~m\AA\ is assumed, which is well within the systematic uncertainty of the RGS. Correcting for this small difference (and implicitly assuming that the \ion{O}{7} line shows no intrinsic redshift), we find an energy of 527.30 $\pm$ 0.05 eV for the \ion{O}{1} K$\alpha$ line. In summary, a list of the \ion{O}{1} $1s\rightarrow 2p$ and $1s-3p$ line energies deduced from astronomical observations is given in Table~\ref{tableenergy}. It must be pointed out here that the listed {\em Chandra} line energies, unlike the {\em XMM-Newton}, were neither re-scaled with the \ion{O}{7} K$\alpha$ line nor Doppler corrected for the motion of the Earth around the Sun. Regarding the former issue, \citet{gatuzz} quotes a line energy of $21.593\pm 0.002$~\AA\ for the {\em Chandra} observation of \ion{O}{7} K$\alpha$ towards XTE~J1817-330 which is 9~m\AA\ short of the laboratory standard. If the wavelength scale is adjusted accordingly, our {\em Chandra} \ion{O}{1} K$\alpha$ position in Table~\ref{tableenergy} would be reduced to $527.26\pm 0.09$~eV, in much better agreement with the {\em XMM-Newton} value.

Very recently, a new and independent investigation of 36 Chandra HETG observations of 11 low-mass X-ray binaries has appeared, which accounts for the Galactic rotation velocity relative to the rest frame and uses a similar merging of corrected spectra \citep{liao}.  A Bayesian analysis is employed to 
quantify systematic uncertainties  and bias corrections, obtaining a resonance position with improved statistics. The resulting energy position of $527.39\pm 0.02$ eV, as listed in Table~\ref{tableenergy}, is in excellent agreement with our average observed value of $527.37$ eV. 

\subsubsection{Resonance Energy Positions from Laboratory Measurements}
\label{seclabexp}

We now consider the laboratory data for atomic oxygen. In Fig.~\ref{figdif} we show the differences between the measured
resonance energies for two experiments, namely \citet{menzel} and \citet{stolte}. Firstly, there is a systematic, almost linear change of the energy differences with energy. This may be attributed to small remaining calibration uncertainties in at least one of the two data sets. Furthermore, from the scatter it is seen that the correlation between both data sets is much better than suggested by the formal error bars. This is probably due to the fact that the error bars include a systematic uncertainty that may be nearly the same for all transitions. From a linear regression, we obtain for this energy difference (in eV units): $\Delta E = (36.357 \pm 0.010) - (0.0670 \pm 0.0016)E$ with a scatter of 0.03~eV (much smaller than the nominal uncertainties
of 0.10~eV). Next we compare the measurements of \citet{krause} and \citet{caldwell} with those of \citet{stolte}. In this case, we find no significant slope (best fit $0.0029 \pm 0.0037$) and only a constant offset of $0.444 \pm 0.028$ eV. We conclude that the relative energy scales of \citet{krause}, \citet{caldwell} and \citet{stolte} apparently agree, and that most
likely the energy scale of \citet{menzel} is slightly off. All these data sets, however, show a different offset for their
absolute energy scale.

\subsubsection{Resonance Energy Positions from Large MCHF Calculations}
\label{secmchf}

The $R$-matrix calculations seek to span an indenumerably infinite number of bound, autoionizing, and continuum states of \ion{O}{1} within a single, orthonormal basis of configurations and orbitals, and therefore, it becomes difficult to describe any particular state to a very high degree of accuracy. In the present calculation, we are limited to an active space of up to $n=2$ physical orbitals and $\overline{n}=3$ pseudo-orbitals. However, because the dominant $1s\rightarrow 2p$ transition energy is the source of a rather large ($\approx 0.5{-}0.6$~eV) discrepancy between observations and laboratory experiments,
we can shed further light on the issue by appealing to separate, highly correlated theoretical calculations for the initial and final states. To this end, we have used the sophisticated MCHF atomic structure package \citep{ffmchf} to perform a series
of calculations using separate, large configuration-interaction (CI) expansions, thus increasing the basis size to study the convergence of transition energies and oscillator strengths. Specifically, starting with the initial $1s^22s^22p^4(^3P)$ configuration, we use a basis consisting of all configurations obtainable by any single and/or double promotions from the outer $n=2$ orbitals into the active set of orbitals. For $n_{\rm max}=2$,  the additional $1s^22s2p^5$ and $1s^22p^6$ configurations are taken into account;  for  $n_{\rm max}=3$, configurations such as $1s^22s^22p^33\ell$, $1s^22s2p^43\ell$, $1s^22s^22p^23\ell 3\ell^\prime$, $1s^22s2p^33\ell 3\ell^\prime$, and $1s^22p^23\ell 3\ell^\prime$ ($3\ell=\left\{3s,3p,3d \right\}$) are also included.  This procedure is repeated for $n_{\rm max}=4,5,6$, and at each stage, a full-scale MCHF calculation is performed, optimizing each of the separate orbitals from $n=1$ to $n=n_{\rm max}$ to produce a lengthy multi-configuration wavefunction for the initial state. This same procedure is likewise repeated for the final state, re-optimizing all of the orbitals separately. Lastly, for given initial and final wavefunctions (using completely different, non-orthogonal orbital bases), the absolute energies, transition energies, and oscillator strengths are computed. The results are listed in Table~\ref{tablemchf} where it is interesting to see how the transition energy oscillates significantly between the observed and experimental values at first, but converges to a value consistent with that determined from the X-ray observations.

\subsubsection{Final Resonance Fit}
\label{secfinalfit}

\begin{figure}[!hbtp]
\centering
\includegraphics[width=2.5in,angle=-90.]{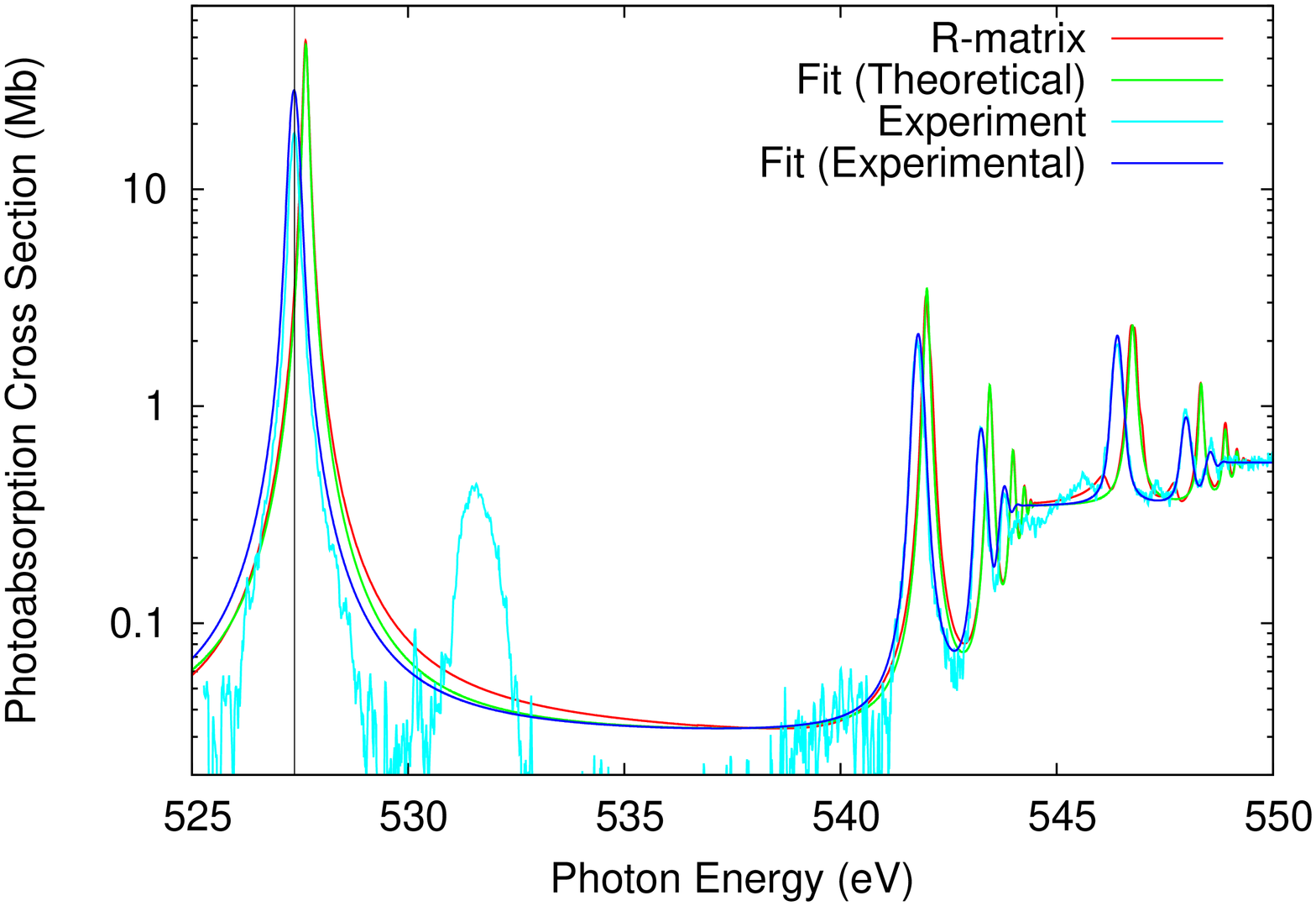}
\includegraphics[width=2.5in,angle=-90.]{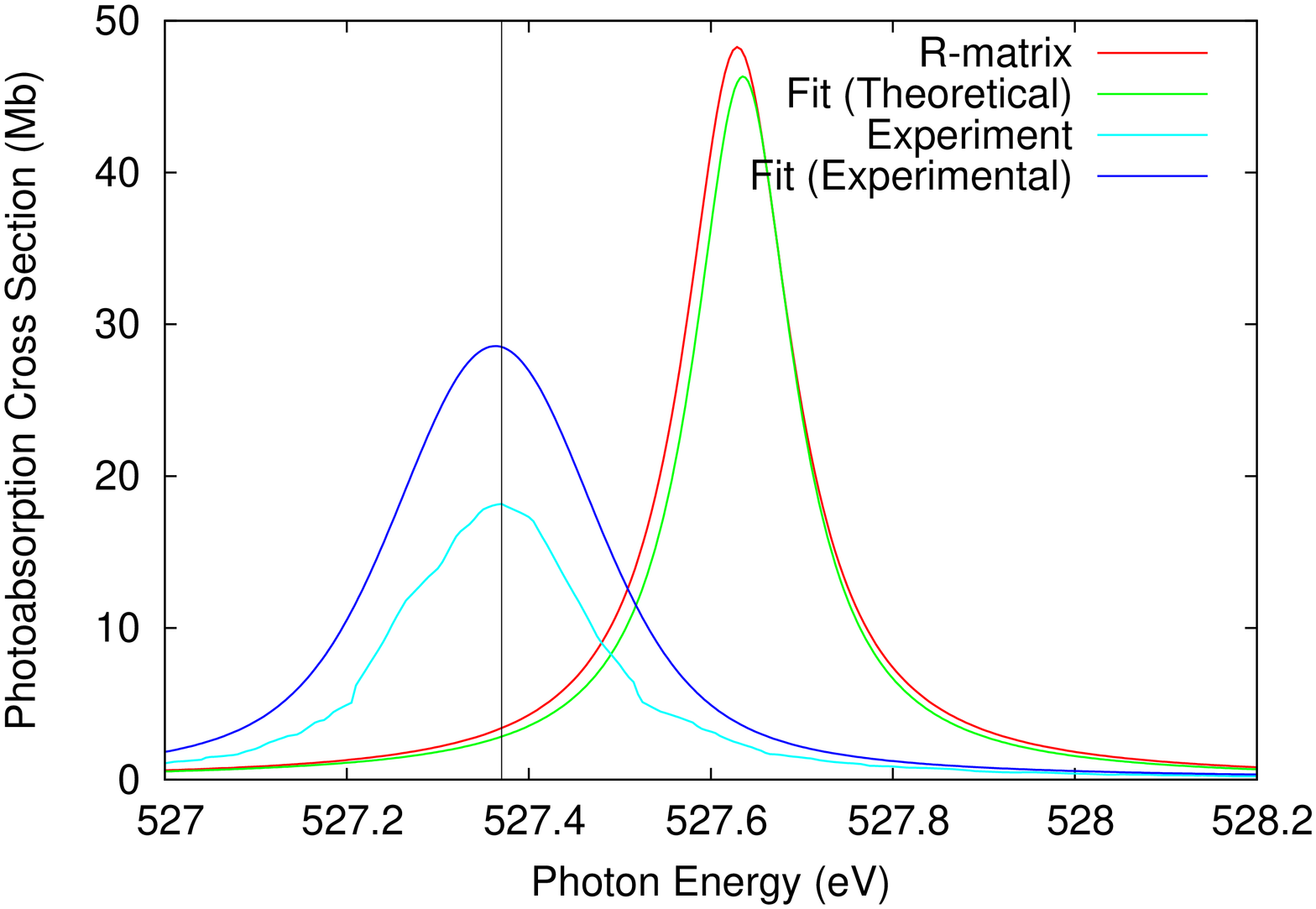}
\includegraphics[width=2.5in,angle=-90.]{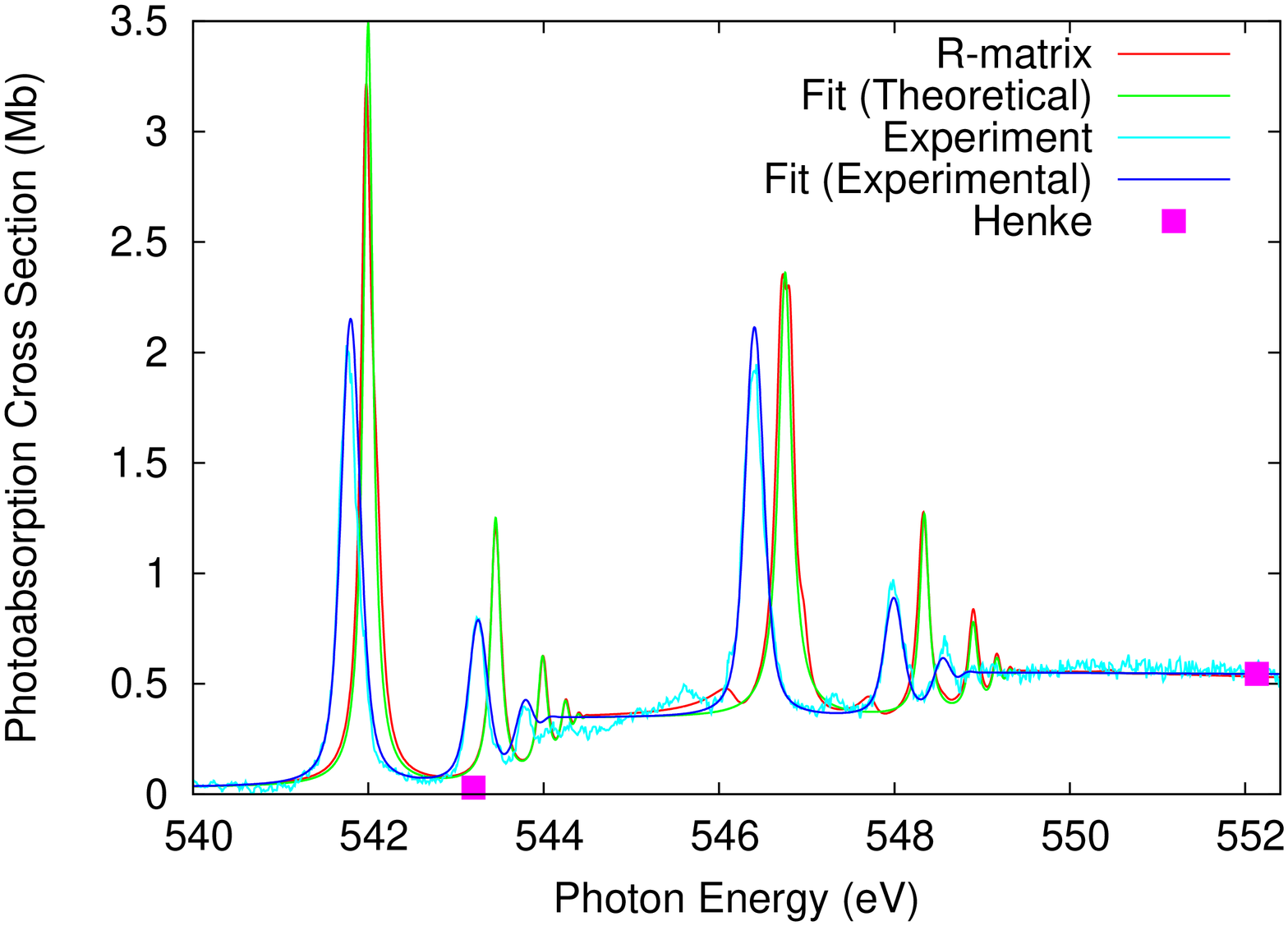}
\caption{\label{figexp}
Fitting comparison:
{\it R}-matrix results (red curve),
analytical fit formula to theoretical results (green curve),
experimental results of \citet{stolte}, shifted by +0.58 eV (cyan data points),
analytical fit formula using adjusted threshold energies and convoluted with the experimental resolution of 182 meV (cyan curve), \citet{henke} data (magenta squares),
 $1s\rightarrow 2p$ resonance position of 527.37 eV determined from observation (black vertical line, see  text).
}
\end{figure}

By fitting the expression in Eq.~\ref{eqfit} to the $R$-matrix results, we obtain the parameters that are listed in Table~\ref{tablefit}. However, our initial fit used---in addition to the same widths as determined in the $R$-matrix run
(0.1348~eV and  0.1235~eV for series $i_s=1$ and $i_s=2$, respectively)---the theoretical threshold energy positions $E_{th}^{1}=544.74$ eV and $E_{th}^{2}=549.67$ eV. As is seen at all levels in Fig.~\ref{figexp}, this prescription provides an excellent fit to the $R$-matrix results. The fit formula was next compared to the experimental cross section of
\citet{stolte} (shifted by +0.58 eV in order to position the $1s\rightarrow 2p$ resonance at 527.37~eV). However, in order to align our fit with these shifted experimental results, which comprise our assessment of the most accurate resonance positions, we had to shift our theoretical threshold energies by $-0.2$~eV and $-0.35$~eV for the two series $i_s=1$ and $i_s=2$, respectively. This has the simple effect of shifting every individual resonance of each series by these amounts. Furthermore, in order to obtain the most meaningful comparison, it was necessary to convolute the fitting expression with a Gaussian of 182~meV FWHM to simulate the experimental resolution. Finally, it was necessary to upscale the $n=2$ oscillator strength to match the more reliable MCHF value of 0.097 (see Table~\ref{tablemchf}); the R-matrix $n=2$ oscillator strength was scaled down by a factor of $0.80$ from the $n\rightarrow \infty$ series limit, and we use as the final fit the slightly increased value $f_{0,2}^1=0.867f_{0,\infty}^1$, which ensures an oscillator strength of
$f=f^1_{0,2}/(2-\mu^1_2)^3 =0.097$.

With these final adjustments, the resulting fit is seen to reproduce the experimental results very well except for the $n=2$ resonance strength. Note that in particular the quantum defects for each of the two series, as determined from the fit to the $R$-matrix results, align satisfactorily with the experimental values indicating that, regarding energy determination, the main source of error lies in determining accurate threshold positions. Furthermore, with respect to the experimental spectrum, it seems that, if there is any error at all, it must be an overall global offset that we assume here to be $-0.58$~eV. We note that the fit formula of Eq.~\ref{eqfit} could have just been modified by replacing the unit Lorentzian profiles by unit Voigt profiles, but it is actually straightforward to simply convolute the resulting cross section numerically. Nevertheless, this highlights the flexibility of the analytical fitting formula in that the particular resonance shape can be, if desired, easily accounted for analytically.

As a last key point, we comment that experiment has a noticeable signal due to molecular O$_2$ contamination in the beam as evidenced by the strong $1s\rightarrow \pi^*$ resonance at 531~eV.  Consequently, there is an additional signal in the experiment throughout the K-edge region, and the experimental procedure does not provide the most accurate benchmark away from resonance. On resonance, the signal is so strong and predominantly atomic in nature that the small molecular admixture would not affect the oscillator strength as much.

\section{Comparison to Other Models}

The three data sets of X-ray absorption currently used in spectral modeling codes we wish to compare to the present fit are: (1) the $R$-matrix cross sections of \citet{o}; (2) the {\sc xstar} modeling code which uses the $R$-matrix cross sections of \citet{Garcia} (except for independent data for the the lowest $1s\rightarrow 2p$ resonance); and (3) {\sc spex} which is based on calculations with {\sc hfr} \citep{cowan} for resonances and data by \citet{verner96} elsewhere. A comparison is given in Fig.~\ref{figcomp} between these various approaches and the present fit where several points must be brought to light. First, the cross section of \citet{o}, at least with respect to resonance positions, is close to the present fit since it is based on similar $R$-matrix calculations. However, the resonance oscillator strengths and the above-threshold cross section are higher, and this was explained in Section \ref{secrmat} as being due to pseudo-resonance contamination present in the earlier $R$-matrix calculations.  Furthermore, the earlier $R$-matrix results show a minor discontinuity at the low-energy tail of the $1s2s^22p^4(^4P)3p$ resonance ($E\approx 538$~eV) due to the sudden turn off of spectator Auger damping which, as discussed in Section  \ref{secrmat}, we alleviate in the present study.

The $R$-matrix results in {\sc xstar} by \citet{Garcia} are seen in Fig.~\ref{figcomp} to be even higher than the present fit, regarding both the resonance oscillator strengths and the above-threshold cross section, and this is believed to be due to an insufficient treatment of both configuration interaction and relaxation effects. In that work, relaxation is partially accounted for by optimizing the orbitals on a weighted sum of closed-$1s$-shell and $1s$-vacancy states of \ion{O}{2} \citep{Garcia}. Lastly, it must be emphasized that the {\sc spex} model has gaps in the total oscillator strength density since only a finite number of terms are included for each Rydberg series. The unconvoluted natural widths are underestimated (in the original data, at least) since the spectator Auger decay was not taken into account; the included participator Auger width approaches zero as $n$ increases ($\Gamma^p_n\sim n^{-3}$). Additionally, the above-threshold cross section is matched to the IP results of \citet{verner96} which, as discussed in Section \ref{secrmat}, underestimates the threshold value.

We also show the newer {\it R}-matrix results of \citet{mclaughlin2}, and they are seen to closely reproduce
the earlier {\it R}-matrix results of \citet{o} including the overestimate of the above-threshold cross section,
as discussed in Section \ref{secrmat}. One additional shortcoming in the new {\it R}-matrix results is that, since spectator Auger decay is not implicitly accounted for in that formulation, the predicted natural widths are underestimated (and, indeed, scale unphysically as $1/n^3$ as $n\rightarrow \infty$).

\section{Summary of Fitting Formula}

The final expression for the photoabsorption cross section consists of the sum of the cross sections in Eq.~\ref{eqtotal}, where
$\sigma_{2s,2p}(E)$, $\sigma^{\rm res}_{1s}(E)$ and $\sigma^{\rm direct}_{1s}(E)$ are given by Eqs.~\ref{eqsig2s2p}, \ref{eqfit}, and \ref{eqfit2}, respectively, and the required fitting parameters are listed in Table~\ref{tablefit}. This final expression has several desirable features which we would like to reinforce.
\begin{itemize}
  \item It is an analytical formula easily transportable between different platforms and modeling codes; in fact, the Fortran routine used to generate a numerical photoabsorption cross section for all energies involves only about one hundred lines of code.
  \item The formulation contains adjustable fitting parameters to best represent: (1) the K-edge positions; (2) the $n\rightarrow \infty$ energy-independent quantum defects and oscillator strengths; and (3) the energy-dependent quantum defects and oscillator strengths for the lower two resonances. From this fit, all relevant atomic parameters can be read off; for instance, the strongest $1s\rightarrow 2p$ oscillator strength can be computed from our fit as
      $$
        f=f^1_{0,2}/(2-\mu^1_2)^3 =0.097\ ,
      $$
      and the integrated resonance strength is therefore given by $\beta f=10.65$ Mb-eV. Further modifications to these parameters can be made if so desired.
  \item The energy spectrum is optimized on the resonance positions determined from a combined experimental and observational assessment.
  \item A constant-resonance-width cross section---a Lorentzian profile that is predicted on physical grounds due to spectator Auger broadening---is implicitly included in the final expression, and can be further modified analytically to include additional broadening effects.
  \item A consistent threshold formulation is obtained in that the $\lim_{n\rightarrow \infty}f_n$ series limit for the (scaled) oscillator strength joins analytically and smoothly to the above-threshold oscillator strength density $df/dE$.
  \item The consistent above-threshold cross section has the  important factor of 0.80 reduction due to relaxation effects, and is then extended to higher energies to include the shake-up and shake-off processes, which result in photoionization-excitation and double photoionization contributions to account for the 20\% difference, and giving the correct $E\rightarrow\infty$ high-energy asymptote, i.e. an accurate ``shoulder".
\end{itemize}

\section{Conclusion}

We have developed an analytical expression that encapsulates all of the important physics in X-ray absorption of atomic oxygen at all photon energies relevant to spectral modeling. For energies below or above the K-edge resonance region, we use simple parametric fits to our best assessment of the cross section based on a convergence between experimental and theoretical data.  The strong $1s\rightarrow np$ resonances belonging to the two dipole-favored Rydberg series, on the other hand, require special attention regarding the oscillator strengths (and analytic continuation to the above-threshold direct $1s$ cross section) and resonance positions. For this important region, we appeal to a combination of $R$-matrix and MCHF theoretical calculations, laboratory experiments, and X-ray astronomical observations. An outstanding issue that we wish to underline is the rather large discrepancy of $\approx 0.6$~eV between several recent observational assessments and the latest laboratory experiments. Unconventionally, we have chosen to use the final calibration as suggested by the observations, since several sources and an independent large MCHF calculation tend to add credibility to this choice. Furthermore, the recent laboratory experiments \citep{stolte,mclaughlin2} calibrated the photon energy scale using the molecular oxygen Rydberg resonance features, and it is unclear to us how accurately those {\em molecular} positions are known, especially considering the uncertainties we find facing us regarding the {\em atomic} resonance positions. We note that a repeat of those experimental measurements, calibrated instead to the more
well-known CO and CO$_2$ K-edge features, will be performed in the near future \citep{wayne2013}, and this will certainly help to shed more light on the 
existing discrepancy.

The ultimate goal of the present work is to establish a definitive, transparent, and easily transportable photoabsorption cross section that can be incorporated in the two spectral modeling codes, namely {\sc xstar} and {\sc spex}. The consistent use of this developed photoabsorption expression in both methods to address molecular abundances in the ISM, and clarify the existing controversy regarding the atomic--molecular fractions \citep{Garciaoxygen}, will be the subject of a subsequent follow-up paper.

\section{Acknowledgments}

TWG acknowledges support by NASA (NNX11AF32G). STM acknowledges support by DOE, Office of Chemical Sciences, Atomic, Molecular and Optical Sciences Program (DE-FG02-03ER15428).

\bibliographystyle{aa}

\begin{figure}[!hbtp]
\centering
\includegraphics[width=3.5in,angle=-90.]{dif.ps}
\caption{\label{figdif}
Comparison between the experimental resonance positions of \citet{menzel} and \citet{stolte}.
}
\end{figure}


\begin{figure}[!hbtp]
\centering
\includegraphics[width=2.5in,angle=-90.]{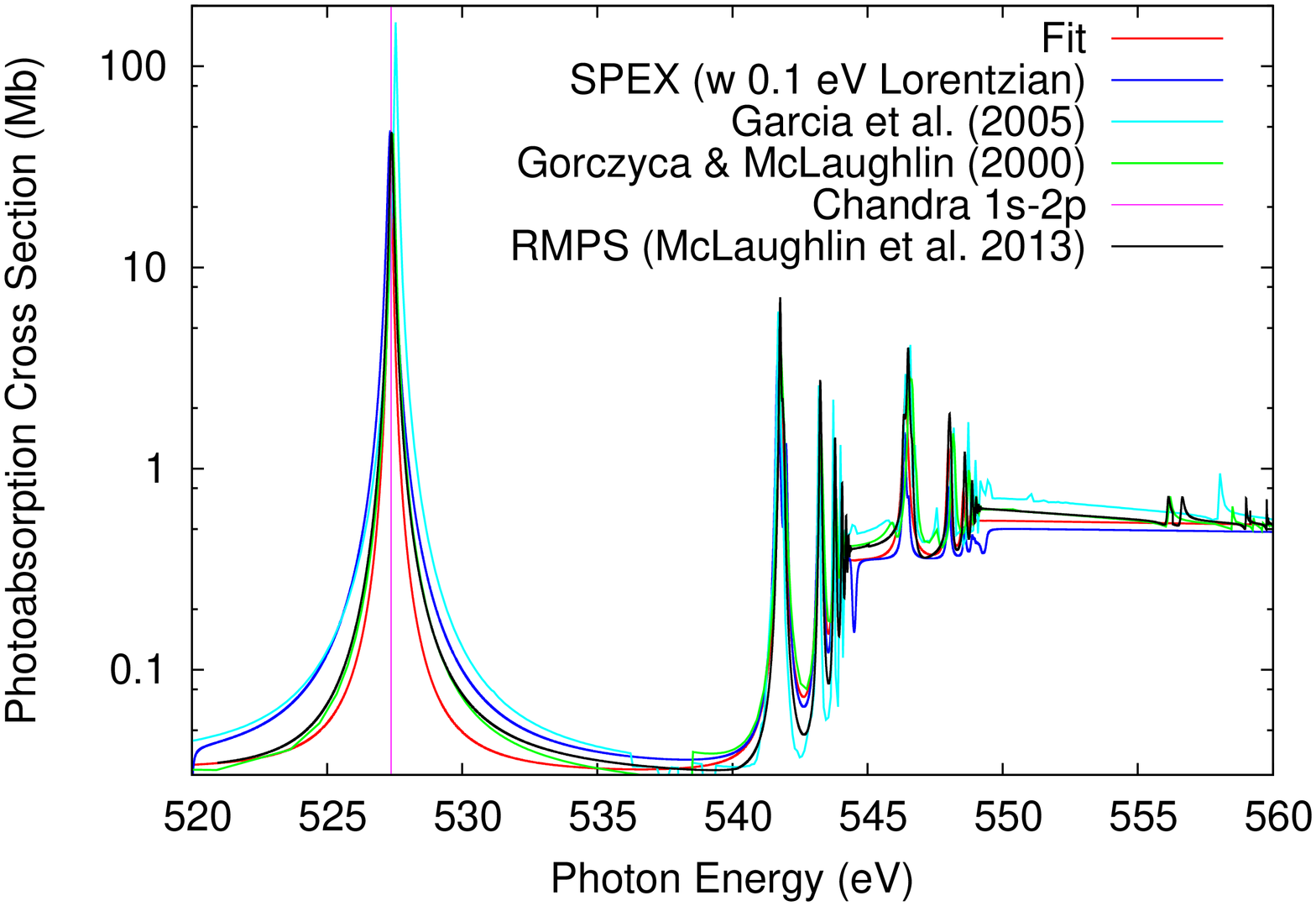}
\includegraphics[width=2.5in,angle=-90.]{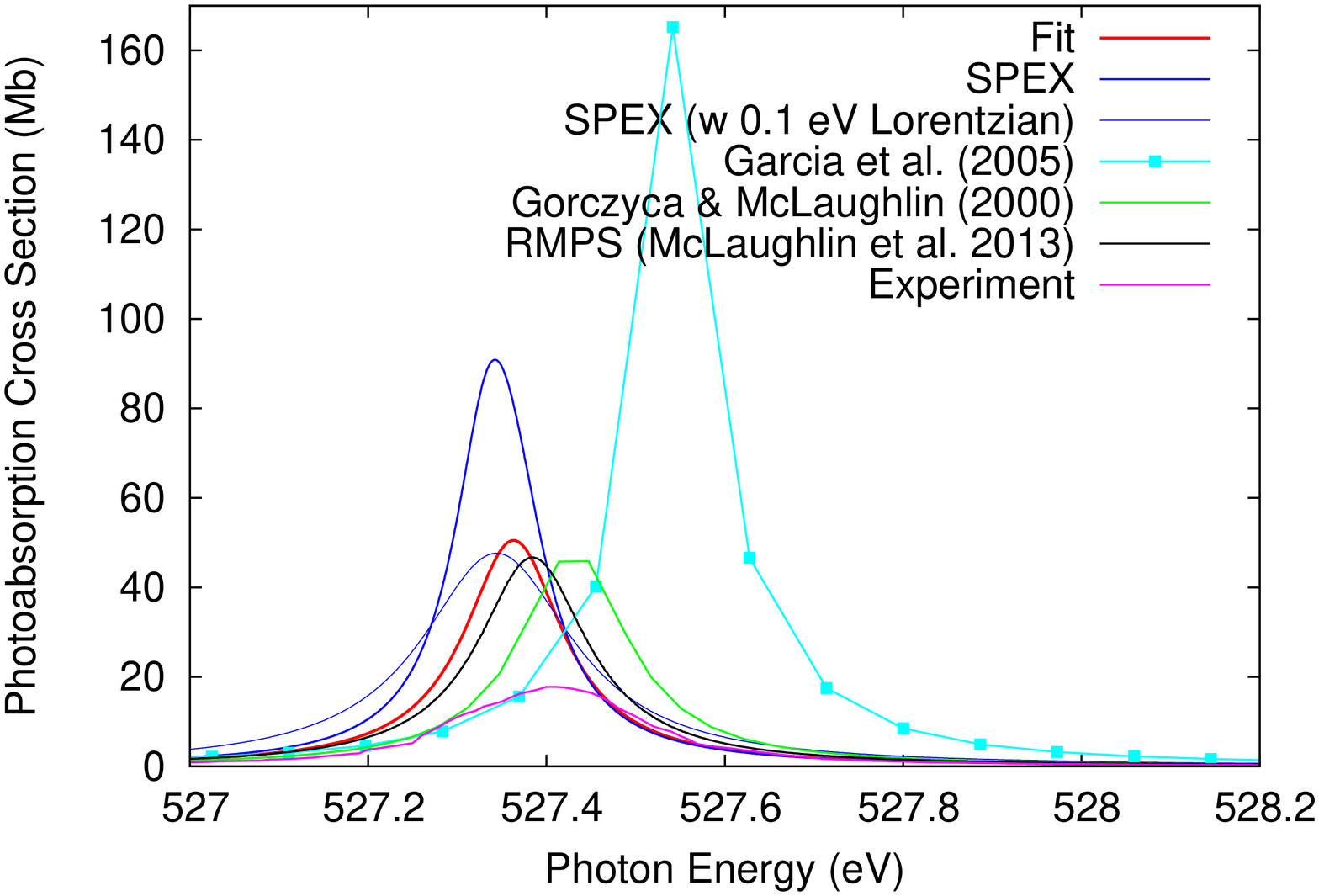}
\includegraphics[width=2.5in,angle=-90.]{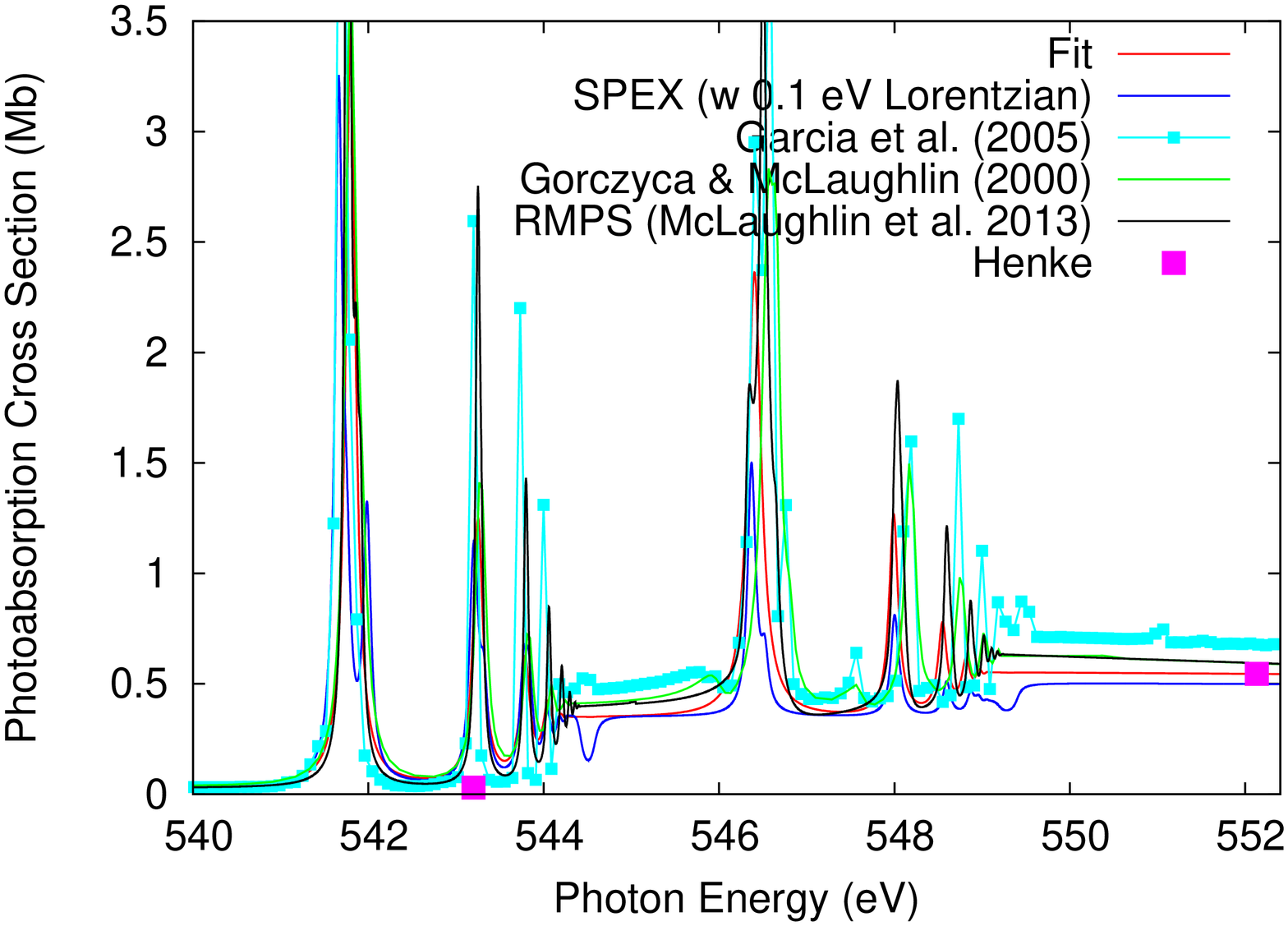}
\caption{\label{figcomp}
Comparison between various data sets:
the present fit model (red curve), SPEX (blue curve), {\it R}-matrix results of \citet{Garcia} (cyan curve), {\it R}-matrix results of \citet{o} (green curve), {\it R}-matrix results of \citet{mclaughlin2} (black curve), \citet{henke} data (magenta squares), present 
Chandra position of the $1s\rightarrow 2p$ resonance at 527.37 eV (magenta vertical line).
The earlier {\it R}-matrix results of \citet{o} and the more recent {\it R}-matrix results
of \citet{mclaughlin2} are indistinguishable except for the lowest
$1s\rightarrow 2p$ resonance in the middle figure.
}
\end{figure}

\newpage

\begin{deluxetable}{ll}
\tablecaption{Summary of Fitting Parameters \label{tablefit}}
\tablewidth{0pt}
\tablehead{
\colhead{Cross section} & \colhead{Parameters}}
\startdata
$\sigma_{2s,2p}$ & $\sigma_0=1745.0$, $E_0=1.24$, $y_a=3.784$, $p=17.64$ \\
                 & $y_w=0.07589$, $y_0=8.698$, $y_1=0.1271$ \\
\hline
$\sigma_{1s}^{\rm res}$ & $f_0=0.132$,\ \  $f^1_{0,\infty}=\frac{3}{5}f_0$,\ \  $f^2_{0,\infty}=\frac{2}{5}f_0$\\
                        & {\bf\boldmath $1s2s^22p^4(^4P)np$\ \  Series ($i_s=1$)}\\
                        & $E^1_{th}=544.54$ eV,\ \  $\Gamma^1=0.1348$ eV \\
                        & $\mu^1_2=1.11$,\ \  $\mu^1_3=0.77$,\ \  $\mu^1_n=0.75$ ($n\ge 4$) \\
                        & $f^1_{0,2}=0.867f^1_{0,\infty}$,\ \  $f^1_{0,3}=0.93f^1_{0,\infty}$,\ \ $f^1_{0,n}=f^1_{0,\infty}$ ($n\ge 4$)\\
                        & {\bf\boldmath $1s2s^22p^4(^2P)np$\ \  Series ($i_s=2$)}\\
                        & $E^2_{th}=549.32$ eV,\ \  $\Gamma^2=0.1235$ eV \\
                        & $\mu^2_3=0.84$,\ \  $\mu^2_n=0.80$ ($n\ge 4$) \\
                        & $f^2_{0,3}=1.02f^2_{0,\infty}$,\ \  $f^2_{0,n}=f^2_{0,\infty}$ ($n\ge 4$) \\
\hline
$\sigma^{\rm direct}_{1s}$ & $\alpha_1=-0.7227$,\ \  $\alpha_2=0.2153$ \\
\enddata
\end{deluxetable}

\begin{deluxetable}{lllcll}
\tabletypesize{\scriptsize}
\tablecaption{{\em Chandra} observations used in this work \label{tab1}}
\tablewidth{0pt}
\tablehead{
\colhead{Source}  &\colhead{ObsID} & \colhead{Date} & \colhead{Exposure (ks)} & \colhead{Read mode} \\
\colhead{ }  &\colhead{ } &\colhead{ } &\colhead{ }&\colhead{ }
}
\startdata
 4U~1636-53&105&1999 Oct 20 & 29 &TIMED \\
& 1939&2001 Mar 28 & 27 &TIMED  \\
&6636&2007 Jul 02 & 26 &CONTINUOUS   \\
&6635&2006 Mar 22 & 23 &	CONTINUOUS    \\
4U~1735-44&704&2006 Jun 09 & 24  &TIMED  \\
&6637&2006 Aug 17 & 25 &	CONTINUOUS   \\
&6638&2007 Mar 15 & 23 &CONTINUOUS   \\
4U~1820-30&1021&2001 Jul 21 & 9.6 &TIMED \\
&1022&2001 Sep 12 & 11 &	TIMED    \\
&6633&2006 Aug 12 & 25 &CONTINUOUS   \\
&7032&2006 Nov 05 & 47 &CONTINUOUS   \\
&6634&2010 Oct 20 & 26 &CONTINUOUS   \\
Cygnus~X-1&3407&2001 Oct 28 & 17 &	CONTINUOUS   \\
&3724&2002 Jul 30 & 8.8 &CONTINUOUS   \\
Cygnus~X-2&1102&2003 Sep 23 & 28  &TIMED  \\
&8599&2007 Aug 23 & 59 &CONTINUOUS  \\
&8170&2007 Aug 25 & 65 &	CONTINUOUS \\
&10881&2009 May 12 & 66 &CONTINUOUS   \\
GX~9+9&703&2000 Aug 22 & 20 &TIMED \\
&11072&2010 Jul 13 & 95 &TIMED \\
XTE~J1817-330&6615&2006 Feb 13&18 &CONTINUOUS \\
&6616&2006 Feb 24&29 &CONTINUOUS \\
&6617&2006 Mar 15&47&CONTINUOUS \\
&6618&2006 May 22&51&CONTINUOUS \\
\enddata
\end{deluxetable}

\begin{deluxetable}{lllll}
\tabletypesize{\scriptsize}
\tablecaption{\ion{O}{1} K$\alpha$ and K$\beta$ line positions (\AA) \label{tab2}}
\tablewidth{0pt}
\tablehead{
\colhead{}    &  \multicolumn{2}{c}{K$\alpha$ ($1s-2p$)} &  \multicolumn{2}{c}{K$\beta$ ($1s-3p$)}\\
\colhead{Source} &\colhead{Present} &\colhead{\citet{juett}} &\colhead{Present} &\colhead{\citet{juett}}
}
\startdata
4U~1636-53&$23.509^{+0.006}_{-0.004}$&$23.507\pm 0.011$&$22.889^{+0.005}_{-0.004}$&$22.915\pm 0.013$\\
4U~1735-44  &$23.507\pm 0.009$&$23.503\pm 0.009$  &$22.890^{+0.009}_{-0.006}$&$22.861\pm 0.006$\\
4U~1820-30  &$23.509\pm 0.004$ &$23.514\pm 0.010$&$22.880^{+0.009}_{-0.010}$&$22.867\pm 0.015$\\
Cygnus~X-1  &$23.507^{+0.003}_{-0.004}$&$23.511\pm 0.007$&$22.882^{+0.011}_{-0.009}$ &$22.888^{+0.023}_{-0.016} $ \\
Cygnus~X-2  &$23.508\pm 0.002$&$23.508\pm 0.004$&$22.883^{+0.010}_{-0.006}$&$22.877^{+0.028}_{-0.026}$  \\
GX~9+9&$23.505^{+0.006}_{-0.004}$&$23.517\pm 0.009$&$22.900^{+0.008}_{-0.010}$&$22.906\pm 0.018$\\
Mean position$^{a}$&$23.507\pm 0.005$&$23.510\pm 0.008$&$22.886 \pm 0.008$&$ 22.885^{+0.017}_{-0.015} $\\
XTE~J1817-330  &$23.506\pm 0.001$&    &$22.889\pm 0.004$& \\
Mean position$^{b}$&$23.507\pm 0.004$&$23.510\pm 0.008$&$22.887 ^{+0.008}_{-0.007 }$&$ 22.885^{+0.017}_{-0.015}  $\\
\enddata
\tablenotetext{a}{Excluding XTE~J1817-330}
\tablenotetext{b}{Including XTE~J1817-330}
\end{deluxetable}

\begin{deluxetable}{llll}
\tablecaption{Line energies (eV) for \ion{O}{1}\label{tableenergy}}
\tablewidth{0pt}
\tablehead{
\colhead{Data set} & \colhead{$1s\rightarrow 2p$} & \colhead{$1s\rightarrow 3p$} & \colhead{$\Delta E$}
}
\startdata
Astronomical observations:          &                  &                  &                \\
{\em Chandra}, average of 7 sources & $527.44\pm 0.09$ & $541.72\pm 0.18$ & $14.28\pm 0.21$\\
{\em XMM-Newton}, Mrk 421           & $527.30\pm 0.05$ & $541.95\pm 0.28$ & $14.65\pm 0.33$\\
\citet{juett}, 6 sources            & $527.41\pm 0.18$ & $541.77\pm 0.40$ & $14.36\pm 0.58$\\
{\bf Average}                       & $\bf 527.37$     &                  &                \\
{\em Chandra}, \citet{liao}                       & $527.39\pm 0.02$     &                  &                \\
\hline
Laboratory measurements:            &                  &                  &                \\
\citet{mclaughlin2}                 & $526.79\pm 0.04$ & $541.19\pm 0.04$ & $14.40\pm 0.08$ \\
\citet{stolte}                      & $526.79\pm 0.04$ & $541.20\pm 0.04$ & $14.41\pm 0.08$ \\
\citet{krause},                     &                  &                  &                 \\
\ \ \citet{caldwell}                & $527.20 \pm 0.30$&                  &                 \\
\citet{menzel}                      & $527.85 \pm 0.10$& $541.27 \pm 0.15$& $13.41\pm 0.25$ \\
\hline
MCHF calculations:                  &                  &                  &                 \\
($n_{\rm max}=6$)                   & 527.49           &                  &                 \\
\enddata
\end{deluxetable}

\begin{deluxetable}{cccccc}
\tablecaption{MCHF data for $1s^22s^22p^4(^3P)\rightarrow 1s2s^22p^5(^3P)$ in \ion{O}{1} \label{tablemchf}}
\tablewidth{0pt}
\tablehead{
\colhead{$n_{\rm max}$} & \colhead{$E_i$} & \colhead{$E_f$} & \colhead{$\Delta_E$} & \colhead{$f_L$} & \colhead{$f_V$}\\
\colhead{} & \colhead{(a.u.)} & \colhead{(a.u.)} & \colhead{(eV)} & \colhead{} & \colhead{}
}
\startdata
2 & -74.85830 & -55.44337 & 528.29  & 0.133 & 0.121 \\
3 & -74.99720 & -55.63645 & 526.82  & 0.107 & 0.102 \\
4 & -75.06477 & -55.68599 & 527.31  & 0.098 & 0.101 \\
5 & -75.08774 & -55.70510 & 527.41  & 0.093 & 0.097 \\
6 & -75.09707 & -55.71152 & 527.49  & 0.097 & 0.096 \\
\enddata
\tablecomments{Results are given as a function of $n_{\rm max}$, the maximum principal quantum number included in the active space expansion of configurations obtained by single and double promotions out of the initial or final configuration. Separate orbital bases are used for initial and final states, and relativistic corrections account for an additional $\approx 0.03$ eV to the transition energy. $f_L$ and $f_V$ are respectively the oscillator strengths in the length and velocity gauges.}
\end{deluxetable}

\end{document}